\documentclass[english]{sbrt}

\usepackage{physics}
\usepackage{amssymb, amsmath,amsfonts,bbm,bm}

\renewcommand{\vu}[1]{\ensuremath{\hat{#1}}}
\renewcommand{\tr}{\ensuremath{\operatorname{tr}}}

\usepackage{subcaption}

\usepackage{pgf}
\usepackage{pgfplots}
\usepackage{pgfplotstable}
\usepackage{tikz}
\usetikzlibrary{plotmarks} 
\pgfplotsset{compat=newest}

\usetikzlibrary{positioning,shapes,shapes.misc,arrows,calc}

\pgfplotsset{graf/.style= {
	thick,
	width=\columnwidth, height = 0.735\columnwidth,
	cycle list = {
	{blue}, {red}, {black}%
	},
    legend pos = north west,
    xtick distance = {0.1},
    ymin=0.01, ymax = 1, 
    xmin=0, xmax = 0.5,
    ylabel= bits/use,
    xlabel= $\tau$}}

\pgfplotsset{markopt1/.style={mark=o}}

\pgfplotsset{markopt2/.style={mark=x}}

\pgfplotsset{const/.style= {
    width=\columnwidth, height =\columnwidth,
    axis x line=center, axis y line=center, axis equal,
    ticks=none, xlabel={$q$}, ylabel={$p$},
    xmin=-2, xmax=2, ymin=-2, ymax=2,
    }}

\pgfplotsset{graf2/.style= {
	width=\columnwidth,
	cycle list = {
	{blue}, {red}, {black}, {brown}%
	},
    legend pos = south east,
    no markers, 
    xtick distance = {40},
    ymin=0.01, ymax = 2, 
    xmin=0, xmax = 200,
    ylabel= bits/use,
    xlabel= Distance (km)}}

\usepackage{cleveref}

\usepackage[bibstyle=ieee,citestyle=numeric-comp,sorting=none,backend=bibtex]{biblatex}
\begin{document}
\title{Amplitude-Phase Modulated CVQKD Protocol}

\author{Micael Andrade Dias and Francisco Marcos de Assis
\thanks{Micael Andrade Dias and Francisco Marcos de Assis, Center of Electrical Engineering and Informatics, Federal University of Campina Grande (UFCG), Campina Grande-PB, Brazil. E-mails: micael.souza@ee.ufcg.edu.br, fmarcos@dee.ufcg.edu.br. This work was partially supported by CAPES.}%
}

\maketitle

\markboth{XXXIX SIMP\'{O}SIO BRASILEIRO DE TELECOMUNICA\c{C}\~{O}ES E PROCESSAMENTO DE SINAIS - SBrT 2021, 26--29 DE SETEMBRO DE 2021, FORTALEZA, CE}{}

\begin{abstract}
	Here we report the performance of a discrete modulated Continuous Variable Quantum Key Distribution protocol based on eight and sixteen state optimal Amplitude Phase Keying constellations with pure loss quantum channel transmission. Our constellations use the same amplitudes and state probability as in an optimal four-state unidimensional constellation model and apply amplitude/phase modulation to the quadratures of coherent states. We compare the obtained results with the performance of conventional $m$-PSK and $m$-APK constellations. The results presented show that protocols that use the proposed constellations outperform the ones based on conventional constellations in both secret key rate and link distance. In special, our protocol is able to accomplish viable SKR's in low transmittance regime, $\tau<0.2$.

\end{abstract}
\begin{keywords}
CVQKD, Discrete Modulation, Constellation Shaping.
\end{keywords}

\section{Introduction}\label{sec:intro}

Quantum Key Distribution (QKD) protocols, one of the most promising Quantum Information practical implementations, are mainly divided into protocols based on discrete variable quantum systems, represented by finite dimensional Hilbert spaces, and on continuous variable quantum systems with infinite dimensional Hilbert spaces \cite{weedbrook2012}. The first class is known as Discrete Variable QKD (DVQKD) protocols and naturally applies discrete modulation to encode information as in, for example, the BB84 protocol with single photon polarization and detection \cite{bennett2014}. These protocols are extremely sensitive during quantum communication phase but allows the use of simple error correction schemes such as CASCADE protocol \cite{brassard1994}. The QKD protocols with continuous variables (CVQKD protocols) usually use coherent or squeezed quantum states and, at first, applied discrete modulation \cite{ralph1999} (finite set of transmitted quantum states) but, with the GG02 protocol \cite{grosshans2002}, Gaussian modulation of quadratures became a reference system for CVQKD. These continuous variable based protocols are, in principle, compatible with off-the-shelf optical networks devices (modulators, detectors, multiplexers, etc.) but demands well-designed continuous-to-discrete operations and state-of-the-art capacity achieving error correcting codes \cite{assche2004,nguyen2004,bloch2006,bai2017,jouguet2014}.

In the last decade, research on QKD has turned to a hybrid solution, intending to harvest the best characteristics of both families, by applying discrete modulation on continuous variable quantum systems (DM-CVQKD). With these protocols, there is no need for a continuous-to-discrete conversion and the compatibility with current optical network devices is maintained by using coherent states. This conversion from continuous (analog) to discrete (digital) modulation is a process already experienced in classical communication systems where Gaussian modulation, which matches the statistical behavior of an Aditive White Gaussian Noise (AWGN) channel and reaches the channel capacity, is approximated by digital signaling schemes such as Phase Shift Keying (PSK), Amplitude Shift Keying (ASK), Amplitude Phase Keying (APK) and Quadrature Amplitude Modulation (QAM), also known as constellations \cite{proakis2008}.

Although digitally modulated protocols come as a practical solution for deploying QKD, a price is payed as non-Gaussian modulation is suboptimal for secret key rates \cite{wolf2006}. Although the ultimate secret key rate is reached theoretically by Gaussian modulation on both quadratures, several DM-CVQKD protocols were proposed based on $m$-PSK constellations, each one presenting different security proofs and the conditions for which the secrete key rate is close to a Gaussian modulated protocol \cite{zhao2009,leverrier2009,bradler2018,papanastasiou2018,ghorai2019}. 

In the classical context, some measures can be taken to reduce the effects of using digital modulation. Firstly, one can increase the number of states in the signal set (the constellation). A more sophisticated solution is to adjust the constellation shape and assign different probabilities to the states, which is known as geometric and probabilistic shaping, respectively. Several constellation shapes of unidimensional (UD) constellations were analyzed in \cite{wu2010} and it was shown that the ones applying some kind of shaping have a better convergence towards Gaussian modulation performance as its cardinalities increase than the ones with no geometric and/or probabilistic shaping. This capability of approximating Gaussian modulation by shaping the constellation is also translated to QKD protocols, as it was shown in \cite{dias2021} where the authors used the UD constellations in \cite{wu2010} and compared with an UD Gaussian modulation \cite{usenko2015}, and they concluded that capacity archiving\footnote{By capacity achieving the authors mean a modulation scheme whose performance in the AWGN channel reaches the information-theoretic limit of continuous Gaussian modulation.} UD constellations also reaches the performance of Gaussian modulated UD protocol. The authors also presented optimal four state constellations obtained by an exhaustive search on its geometry and probability for each transmittance in the interval $\tau\in\left]0,1\right[$.

Despite the good performance of digital UD modulation, finding bidimensional efficient constellations is still quite challenging.  In this paper we propose new eight and sixteen-states amplitude-phase shifted constellations based CVQKD protocol. The proposed constellations are based on the optimal UD four-state constellations presented in \cite{dias2021}, where we used the four-state constellation amplitudes and probabilities while applying several rotations to obtain proper eight/sixteen-states constellations. We analyze the protocol performance assuming reverse reconciliation and a noiseless quantum channel. The results indicate that our proposed protocol outperforms standard PSK and APK based protocols in the secret key rate and reachable range. These results imply that the way to obtain better performances for discrete modulated CVQKD protocols goes through managing the constellation geometry and the state probability.

The paper is structured as follows. \Cref{sec:dm-cvqkd} presents the general framework of a discrete modulated QKD protocol with continuous variables. \Cref{sec:opt-prot} introduces the optimized constellations and the main results. In \Cref{sec:conc} we give the final considerations and perspectives for future work.

\section{Discrete Modulated Protocols}\label{sec:dm-cvqkd}

Any QKD protocol consists on two legitimate parties (Alice and Bob) intending to establish a secret random stream of symbols to be used as cryptographic key and, for that, they use quantum systems to transmit classical information. This phase of preparing, transmitting and detecting quantum states is called the protocol's quantum stage and is where most differences between many kinds of QKD protocols lie. In a Gaussian Modulated Coherent State (GMCS) protocol, Alice will draw realizations $q_i$ and $p_i$ of normally distributed random variables $\mathcal{Q}$ and $\mathcal{P}$ to define the complex amplitude $\alpha_i = q_i + ip_i$ of a coherent state, which will be prepared, transmitted and detected by Bob with homodyne or heterodyne setups. After state preparation/detection, the complex amplitudes $\alpha_i$ and detection outcomes $\beta_i$ are continuous valued and then require some quantization process in order to result on binary strings. This quantizer must be well designed and works alongside with the error correction protocol. This continuous-to-discrete operation added to error correction provides a bottleneck on the protocol's performance.

\begin{figure}[!bt]
	\centering
	\includegraphics{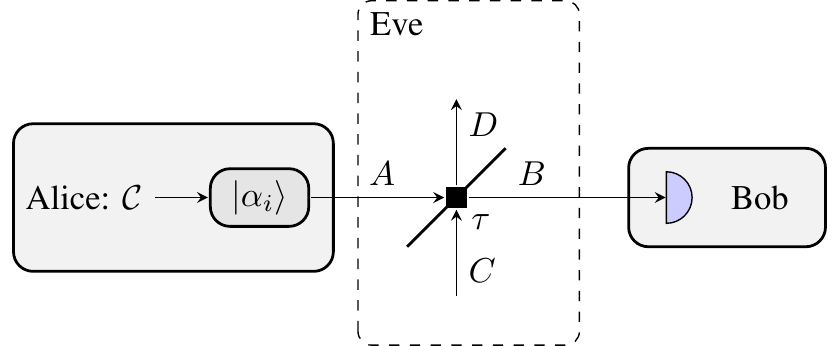}
	\caption{\label{fig:pl-chan}Pure loss quantum channel model for a DM-CVQKD protocol.}	
\end{figure}

A discrete modulated CVQKD protocol dismiss the quantization process and works as follows. Instead of drawing realizations out of normal random variables, define an alphabet of $N = 2^K$ letters to be drawn randomly by Alice according to some suitable probability distribution $\mathcal{P} = \qty{p_i}$, each letter corresponding to a binary word of $K$ bits, and a set of complex amplitudes $\mathcal{A}$ with $|\mathcal{A}| = N$. Each letter is mapped unambiguously to a coherent state $\ket{\alpha_i}$ whose complex amplitude $\alpha_i\in\mathcal{A}$ and we call $C(\mathcal{A},\mathcal{P})$ a constellation \textit{geometric shaped} by $\mathcal{A}$ and \textit{probabilistic shaped} by $\mathcal{P}$. Different constellations define distinct DM-CVQKD protocols although they may share the same classical post-processing protocols\footnote{The security analysis should consider the set of states prepared by Alice in order to estimate the correlations between Alice, Bob and Eve data after quantum state transmission. The protocols in \cite{leverrier2009,zhao2009,bradler2018,ghorai2019} uses constellations with different cardinalities and have its security analysis made from distinct perspectives. Once quantum state transmission is finished, Alice and Bob have correlated binary sequences and, at first, classical post-processing becomes independent of which constellation was used.} (parameter estimation, information reconciliation and security amplification). Then, by randomly drawing the letters, Alice chooses the coherent state that will be prepared on mode $A$, she sends it through a quantum channel whose output mode $B$ will be detected by Bob. We consider the quantum channel to be a pure-loss channel and Bob's detection to be heterodyne, as depicted in \Cref{fig:pl-chan}.

The entire communication (either quantum or classical) is under the presence of a powerful eavesdropper (Eve) whose capabilities are only bounded by the laws of quantum mechanics. Eve will intend to acquire information about the exchanged states by Alice and Bob by performing some information attack during the quantum communication stage\footnote{The eavesdropper attempts during classical error correction of the key post-processing are still considered but the classical channel is assumed to be authenticated, meaning that Eve can not modify the error correction messages or to pass by Alice or Bob.}. This attack comprises the coupling setup, i.e., how she intercept each state sent by Alice, and how she interacts with the states (individual, collective or coherent coupling/detection). In the case of a pure-loss channel, Eve replaces the quantum channel by a beam splitter (BS) with controlled transmittance $\tau$. This BS couples Alice's state to a vacuum state (mode $C$) and one of its outputs ports goes to Bob (mode B) while the other is Eve's (mode $D$). When Eve couples Alice's state to a two-mode squeezed vacuum state, the pure-loss channel becomes a thermal-loss channel and constitutes the entangling cloner attack. 

\subsection{The Secret Key Rate} 
\label{ssub:the_secret_key_rate}

To compute the Secret Key Rate (SKR) of a discrete modulated CVQKD protocol, it is needed to define (or assume) which way to interact with Alice states was chosen Eve. Here we assume the following: (a) pure-loss channel, (b) collective attacks and (c) reverse reconciliation direction. By collective attacks we mean that Eve will interact individually with each state but she can store her ancilla mode in a quantum memory to perform a collective delayed measurement. This strategy allows her to use some information of the reconciliation phase to adjust her measurement apparatus so her obtained information can reach the Holevo bound. The reverse reconciliation direction indicates that during error correction (information reconciliation), Alice will adjust her raw key to Bob's one, forcing Eve to relate her states to Bob's outcomes.

Given the above assumptions, the SKR formula for reverse reconciliation under collective attacks is
\begin{equation}
K_{coll} = I(A;B) - \chi(B,E),
\end{equation}
\noindent where $I(A;B)$ is the Shannon mutual information between Alice and Bob's classical random variables and $\chi(B,E)$ is Holevo's bound on Eve's accessible information on Bob's states \cite{holevo1973,nielsen2010,wilde2017a}. To compute the mutual information, one needs to note that Alice's random variable is a discrete one, the set of amplitudes $\mathcal{A}$ and the probabilities assigned to each amplitude, and Bob's variable is continuous and, in the case of heterodyne measurement, one has that the probability of Bob's outcome $b$ given Alice sent $\alpha_i$ is $p(b|\alpha_i) = e^{-|b - \sqrt{\tau}\alpha_i|^2}/\pi$. Then, the mutual information can be computed using the factorization
\begin{equation}
I(A;B) = H(\mathcal{P}) - \int_\mathbb{C} H(A|B=b)\dd[2]{b},
\end{equation}
\noindent where $H(\cdot)$ is the Shannon entropy and the integral is in the complex plane. The Holevo bound gives the maximal classical mutual information between Eve's measurements and Bob's outcomes. This implies that Eve performs the optimal measurement on her states and is computed by
\begin{equation}\label{eq:holevo-bound}
\chi(B,E) = S(\vu\rho_E) - \int_\mathbb{C} p(b)S(\vu\rho_{E|b})\dd[2]{b},
\end{equation}
\noindent where $\vu\rho_E$ is Eve's average state,
\begin{equation}\label{eq:eve-state}
\vu\rho_E = \sum p_i\op*{\sqrt{1-\tau}\alpha_i},
\end{equation}
\noindent and $\vu\rho_{E|b}$ is Eve's average state conditioned to Bob's measurement outcome $b$,
\begin{equation}\label{eq:eve-cond-state}
\vu\rho_{E|b} = \sum p(\alpha_i|b)\op*{\sqrt{1-\tau}\alpha_i}.
\end{equation}	

Here, $S(\cdot)$ is the von Neumann entropy of a quantum state defined as $S(\vu\sigma) = -\tr(\vu\sigma\log\vu\sigma)$. If the state $\vu\sigma$ has a spectral decomposition $\vu\sigma = \sum_k\lambda_k\op{\sigma_k}$, where $\Lambda = \qty{\lambda_k}$ are the eigenvalues of $\vu\sigma$ (non-negative and summing up to $1$) and $\ket{\sigma_k}$ eigenvectors, then $S(\vu\sigma) = H(\Lambda)$. In the \Cref{eq:holevo-bound} where must be computed the entropy for the states in \Cref{eq:eve-state,eq:eve-cond-state}, which are mixtures of nonorthogonal states, its entropy can be computed using the Gram-Schmidt orthogonalization procedure, as explained in the Appendix.


\section{Amplitude-Shifted Constellation Design}\label{sec:opt-prot}

The design of QKD protocols compatible with current communication devices is a crucial step towards practical implementation and the application of discrete modulation to coherent state based protocols provides a solid advance in this direction. Although non-Gaussian modulation is always suboptimal in the SKR, its performance can be improved with well-designed modulations schemes, which implies the search for constellations that maximizes the protocol performance. In the classical context, finding optimal bidimensional constellations is not a simple task and one can only expect that it is no different in a quantum information protocol. 

Four types of unidimensional constellations were examined in \cite{wu2010} for different signal-to-noise ratios and increasing cardinalities. The performance of such constellations in a QKD context were investigated in \cite{dias2021} and they also presented an optimal four-state unidimensional constellation. The search for these optimal four-state constellations was made exhaustively in the following way. Define $\mathcal{C}^{(4)}$ to be the set of all constellations of four coherent states in the form $\qty{\ket{\alpha_1}, \ket{-\alpha_1}, \ket{\alpha_2}, \ket{-\alpha_2}}$, $\alpha_i\in\mathbb{R}$ and $\alpha_2>\alpha_1$, with probabilities $\qty{p_1,p_1,p_2,p_2}$, where $p_1>p_2 = \frac{1}{2} - p_1$. Call $\vu{\rho}_m$ the average state representing an arbitrary ensemble of $\mathcal{C}^{(4)}$ restricted to the energy constrain $\tr(\vu{\rho}_m\vu{n}) = 1$. Let $I(A;B)(\vu{\rho}_m)$ and $\chi(E,B)(\vu{\rho}_m)$ to be Alice and Bob's mutual information and Eve's accessible information, both assuming the protocol used the constellation represented by $\vu{\rho}_m$. Therefore, the maximal SKR for a four-state unidimensional protocol is given by 

\begin{equation}\label{eq:opt-skr}
R_{max}^{(4)} = \max_{\substack{\vu{\rho}_m \in \mathcal{C}^{(4)}\\\tr(\vu{\rho}_m\vu{n}) = 1}}\qty{I(A;B)(\vu{\rho}_m) - \chi(E,B)(\vu{\rho}_m)}.
\end{equation}

\begin{figure}[!t]
	\centering
	\pgfplotstableread{results/opt-4ask-nb1.txt}\resB 
	
	\begin{subfigure}[t]{\columnwidth}
		\centering
		\includegraphics{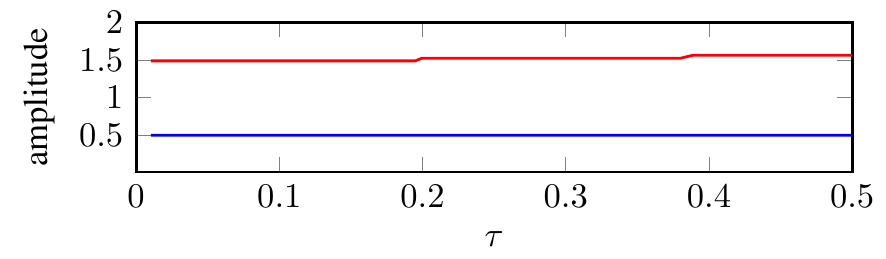}
		\caption{Optimal amplitudes $\alpha_1$ (lower) and $\alpha_2$ (upper).}
	\end{subfigure}
	
	\begin{subfigure}[t]{\columnwidth}
		\centering
		\includegraphics{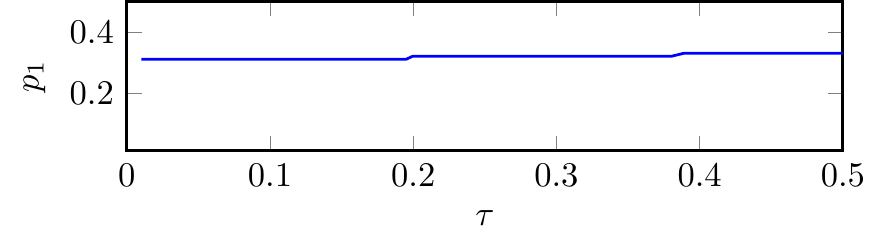}
		\caption{Optimal probability $p_1$.}
	\end{subfigure}
	\caption{\label{fig:opt-alpha}Parameters $\alpha_1$, $\alpha_2$ and $p_1$ for the optimal ensemble that maximizes the SKR in  of a four-state UD-CVQKD protocol.}
\end{figure}

Indeed, by setting the values of $\alpha_1$ and $p_1$ the constellation is completely defined and the search task will look for the pair $(\alpha_1, p_1)$ that maximizes the SKR at a given transmittance $\tau$ with fixed energy constrain. The optimized parameters are replicated in \Cref{fig:opt-alpha} in the interval $0<\tau<0.5$. One can see that, in the presented transmittance range, the amplitude values and the probabilities remained almost unchanged. It can be said that the optimal constellations are slightly geometric shaped (the states are equally spaced) while the probability distribution is not uniform. 

\begin{figure}[!tb]
	\centering
	\begin{subfigure}[t]{0.65\columnwidth}
		\centering
		\includegraphics{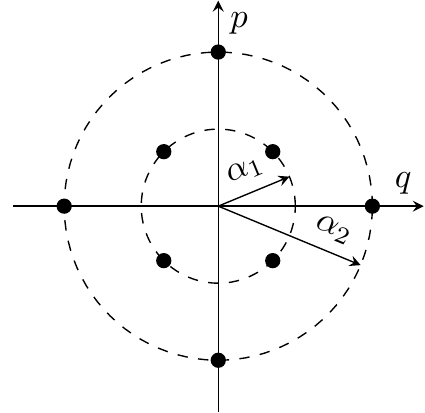}
		\caption{}
	\end{subfigure}
	
	\begin{subfigure}[t]{0.65\columnwidth}
		\centering
		\includegraphics{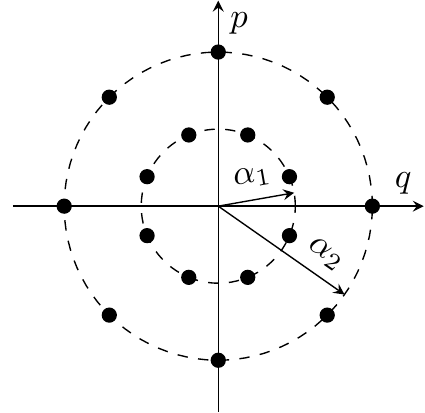}
		\caption{}
	\end{subfigure}
	\caption{\label{fig:const}Eight (a) and sixteen (b) state OAPK constellations designed from the optimal 4UD constellation parameters given by \Cref{eq:opt-skr}.}
\end{figure}

We then used the optimal parameters found by the optimization in \Cref{eq:opt-skr} to design optimal Amplitude-Phase modulations (OAPK) schemes with eight and sixteen states. We denote by $N_1/N_2$OAPK the Optimal Amplitude Phase Keying constellation with $N_1$ states of energy $\alpha_1^2$ and $N_2$ states of energy $\alpha_2^2$. From the probabilities $\qty{p_1,p_2}$ we define that the state with energy $\alpha_i^2$ is prepared with probability $q_i = 2p_i/N_i$. Such constellations are shown in \Cref{fig:const}.

\begin{figure}[!tb]
	\centering
	\includegraphics{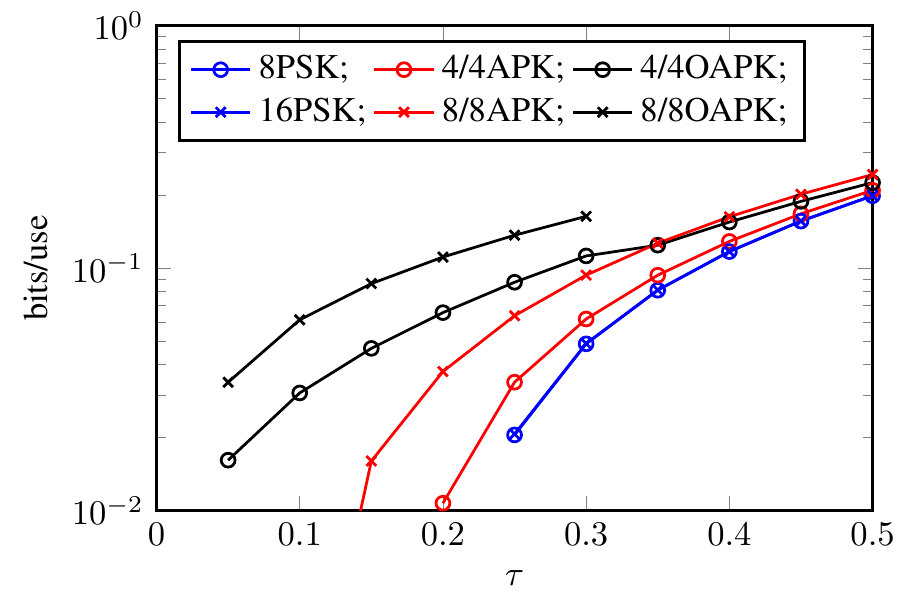}
	\caption{\label{fig:skr} Secret key rates for the discrete modulated CVQKD protocol based on PSK, APK and our proposed optimized APK constellations with eight (circle marker) and sixteen (cross marker) states. All constellations have fixed unit average energy. }
\end{figure}

In \Cref{fig:skr} we plot the SKR for the eight and sixteen state OAPK constellations together with conventional 8PSK, 16PSK, 4/4APK and 8/8APK constellations. At first, the results show that the proposed constellations applied to QKD outperform PSK and APK based protocols. It can be seen that our eight-state protocol provides greater SKR in the entire transmittance range and reaches longer distances (positive SKR's for low transmittance) than the other eight-state protocols. However, the 4/4OAPK protocol performance gets a little diminished for $\tau>0.3$ and we can address this issue to the fact that as the transmittance increases, the states received by Bob becomes more distinguishable by him and the eavesdropper is able to retrieve more information about his states. 

When we turn to the proposed sixteen-state protocol, two things call attention. First, increasing the cardinality of our OAPK protocol considerably increases the secret key rate and, second, there is no significant difference between the 8PSK and 16PSK constellations. The 16APK constellation presents a performance enhancement but still remains undermined by both 4/4OAPK and 8/8OAPK constellations. One can conclude that $m$-PSK based protocols have a saturation point in its cardinality near to eight states. By comparing the traditional APK and our proposed OAPK constellations, we can conclude that simple adjustments to the constellation geometry and probability distribution results in considerably SKR increasing and allows the protocol to reach longer distances. The 8/8OAPK constellation results for $\tau>0.3$ where not presented due to numerical issues.

\section{Conclusion}\label{sec:conc}

We have presented amplitude-phase shifted constellations obtained from unidimensional four-state constellations optimized to CVQKD and analyzed its performance in the reverse reconciliation regime over a pure-loss quantum channel. We computed the secret key rates for the proposed amplitude-phase constellations with eight and sixteen states and compared with conventional $m$-PSK and $m$-APK constellations. The results presented a superiority of the proposed constellations over the conventional ones, providing greater secret key rates and capability of reaching longer distances. Such enhancement shows that simple constellation shaping can improve the protocol's performance. Future works can be focused on finding constellations with greater cardinalities and also including the thermal noise effect in the quantum channel.

\appendix

\section{Ensemble Diagonalization via Gram-Schmidt Orthogonalization Procedure}\label{sec:ap_GS}

Let us define an ensemble of arbitrary coherent states $\mathcal{S} = \qty{\ket{\alpha_i}, p_i}_{i=0}^{K-1}$, being $\alpha_i \in \mathbb{C}$ and $p_i$ the probability of the $i$-th state on $\mathcal{S}$ to be prepared, with $\sum p_i = 1$. From $\mathcal{S}$, one can compute the inner product matrix $\bm{V}_\mathcal{S}$,
\begin{equation}
V_{ij} = \braket{\alpha_i}{\alpha_j} = e^{\frac{1}{2}(\alpha_i^*\alpha_j - \alpha_i\alpha_j^*)}e^{-\frac{1}{2}\abs{\alpha_i - \alpha_j}^2}.
\end{equation}

By the Gram-Schmidt orthogonalization procedure it is possible to compute, using the states on $\mathcal{S}$, an orthonormal set of states $\mathcal{B} = \qty{\ket{\psi_0}, \cdots, \ket{\psi_{K-1}}}$ which is a base to the subspace spanned by $\qty{\ket{\alpha_i}}_{i=0}^{K-1}$. Then, one can decompose each $\ket{\alpha_i}$ with $\mathcal{B}$ basis:
\begin{equation}\label{eq:est_coh_projetado}
\ket{\alpha_k} = \sum_{i = 0}^{K-1}\braket{\psi_i}{\alpha_k}\ket{\psi_i} = \sum_{i = 0}^{K-1}M_{ki}\ket{\psi_i},
\end{equation}
\noindent
where $M_{k,i}$ corresponds to the projection of the $k$-th state $\ket{\alpha_k}$ into the $i$-th basis element of $\mathcal{B}$. These projections are arranged as a matrix, for which each element may be computed by the following algorithm:
\begin{align}
M_{k 0} &= V_{0 k}\\
M_{k i} &= \frac{1}{M_{i i}}(V_{i k} - \sum_{j=0}^{i-1} M_{i j}^{*} M_{k j})\quad if\; 1 \leq i < k\\
M_{k i} &= 0\quad if\; i > k\\
M_{k k} & =\qty(1-\sum_{i=0}^{k-1}|M_{k i}|^{2})^{1/2}\quad for\; k>0.
\end{align}

By the description on \Cref{eq:est_coh_projetado}, the density operators $\vu\rho_{\alpha_k}$ and $\vu\rho_m$ for the coherent state $\ket{\alpha_k}$ and the whole ensemble $\mathcal{S}$ are, respectively,
\begin{align}
\vu\rho_{\alpha_k} &= \sum_{i,j=0}^kM_{ki}M^*_{kj}\op{\psi_i}{\psi_j}\\\label{eq:densidade_ensemble}
\vu\rho_m &= \sum_{k = 0}^{N-1}p_k\vu\rho_{\alpha_k} = \sum_{k = 0}^{N-1}p_k\sum_{i,j=0}^kM_{ki}M^*_{kj}\op{\psi_i}{\psi_j}.
\end{align}

\printbibliography

@article{dias2021,
  title = {The {{Impact}} of {{Constellation Cardinality}} on {{Discrete Unidimensional CVQKD Protocols}}},
  author = {Dias, Micael and Assis, Francisco},
  year = {2021},
  pubstate = {Manuscript submitted for publication},
  journal = {Quantum Information Processing}
}

@article{assche2004,
  title = {Reconciliation of a Quantum-Distributed {{Gaussian}} Key},
  author = {Assche, Gilles Van and Cardinal, J. and Cerf, Nicolas J.},
  year = {2004},
  volume = {50},
  pages = {394--400},
  issn = {0018-9448},
  doi = {10.1109/tit.2003.822618},
  journal = {IEEE TIT},
  number = {2}
}

@article{bai2017,
  title = {High-Efficiency Reconciliation for Continuous Variable Quantum Key Distribution},
  author = {Bai, Zengliang and Yang, Shenshen and Li, Yongmin},
  year = {2017},
  month = mar,
  volume = {56},
  pages = {44401},
  publisher = {{Japan Society of Applied Physics}},
  doi = {10.7567/jjap.56.044401},
  journal = {Japanese Journal of Applied Physics},
  number = {4}
}

@article{bennett2014,
  title = {Quantum Cryptography: {{Public}} Key Distribution and Coin Tossing},
  author = {Bennett, Charles H. and Brassard, Gilles},
  year = {2014},
  volume = {560},
  pages = {7--11},
  publisher = {{Elsevier \{BV\}}},
  doi = {10.1016/j.tcs.2014.05.025},
  journal = {Theoretical Computer Science}
}

@article{bradler2018,
  title = {Security Proof of Continuous-Variable Quantum Key Distribution Using Three Coherent States},
  author = {Br{\'a}dler, Kamil and Weedbrook, Christian},
  year = {2018},
  volume = {97},
  pages = {1--16},
  issn = {24699934},
  doi = {10.1103/PhysRevA.97.022310},
  journal = {Phys. Rev. A},
  number = {2}
}

@inproceedings{brassard1994,
  title = {Secret-{{Key Reconciliation}} by {{Public Discussion}}},
  booktitle = {{{EUROCRYPT}}},
  author = {Brassard, Gilles and Salvail, Louis},
  year = {1994},
  pages = {410--423},
  publisher = {{Springer Berlin Heidelberg}},
  address = {{Berlin, Heidelberg}},
  issn = {978-3-540-48285-7},
  doi = {10.1007/3-540-48285-7_35}
}

@article{ghorai2019,
  title = {Asymptotic {{Security}} of {{Continuous}}-{{Variable Quantum Key Distribution}} with a {{Discrete Modulation}}},
  author = {Ghorai, Shouvik and Grangier, Philippe and Diamanti, Eleni and Leverrier, Anthony},
  year = {2019},
  month = jun,
  volume = {9},
  pages = {021059},
  issn = {2160-3308},
  doi = {10.1103/PhysRevX.9.021059},
  journal = {Physical Review X},
  number = {2}
}

@article{grosshans2002,
  title = {Continuous {{Variable Quantum Cryptography Using Coherent States}}},
  author = {Grosshans, Fr{\'e}d{\'e}ric and Grangier, Philippe},
  year = {2002},
  month = jan,
  volume = {88},
  pages = {57902},
  publisher = {{American Physical Society}},
  doi = {10.1103/PhysRevLett.88.057902},
  journal = {Phys. Rev. Lett.},
  number = {5}
}

@article{jouguet2014,
  title = {High-Bit-Rate Continuous-Variable Quantum Key Distribution},
  author = {Jouguet, Paul and Elkouss, David and {Kunz-Jacques}, S{\'e}bastien},
  year = {2014},
  volume = {90},
  pages = {42329},
  publisher = {{American Physical Society}},
  journal = {Phys. Rev. A},
  number = {4}
}

@article{leverrier2009,
  title = {Unconditional Security Proof of Long-Distance Continuous-Variable Quantum Key Distribution with Discrete Modulation},
  author = {Leverrier, Anthony and Grangier, Philippe},
  year = {2009},
  volume = {102},
  issn = {00319007},
  doi = {10.1103/PhysRevLett.102.180504},
  journal = {Phys. Rev. Lett.},
  number = {18}
}

@book{nielsen2010,
  title = {Quantum {{Computation}} and {{Quantum Information}}: 10th {{Anniversary Edition}}},
  author = {Nielsen, M. and Chuang, I. L.},
  year = {2010},
  issn = {00029505},
  doi = {10.1017/CBO9780511976667},
  isbn = {1-107-00217-6},
  journal = {Optics Book}
}

@article{papanastasiou2018,
  title = {Quantum Key Distribution with Phase-Encoded Coherent States: {{Asymptotic}} Security Analysis in Thermal-Loss Channels},
  shorttitle = {Quantum Key Distribution with Phase-Encoded Coherent States},
  author = {Papanastasiou, Panagiotis and Lupo, Cosmo and Weedbrook, Christian and Pirandola, Stefano},
  year = {2018},
  month = jul,
  volume = {98},
  pages = {012340},
  issn = {2469-9926, 2469-9934},
  doi = {10.1103/PhysRevA.98.012340},
  journal = {Physical Review A},
  number = {1}
}

@book{proakis2008,
  title = {Digital Communications},
  author = {Proakis, John G. and Salehi, Masoud},
  year = {2008},
  edition = {5th ed},
  publisher = {{McGraw-Hill}},
  address = {{Boston}},
  isbn = {978-0-07-295716-7},
  lccn = {TK5103.7 .P76 2008}
}

@article{ralph1999,
  title = {Continuous Variable Quantum Cryptography},
  author = {Ralph, T. C.},
  year = {1999},
  volume = {61},
  publisher = {{American Physical Society (\{APS\})}},
  doi = {10.1103/physreva.61.010303},
  journal = {Phys. Rev. A},
  number = {1}
}

@article{usenko2015,
  title = {Unidimensional Continuous-Variable Quantum Key Distribution},
  author = {Usenko, Vladyslav C. and Grosshans, Fr{\'e}d{\'e}ric},
  year = {2015},
  month = dec,
  volume = {92},
  pages = {062337},
  issn = {1050-2947, 1094-1622},
  doi = {10.1103/PhysRevA.92.062337},
  journal = {Physical Review A},
  number = {6}
}

@article{weedbrook2012,
  title = {Gaussian Quantum Information},
  author = {Weedbrook, Christian and Pirandola, Stefano and {Garc{\'i}a-Patr{\'o}n}, Ra{\'u}l and Cerf, Nicolas J. and Ralph, Timothy C. and Shapiro, Jeffrey H. and Lloyd, Seth},
  year = {2012},
  volume = {84},
  pages = {621--669},
  publisher = {{American Physical Society}},
  journal = {Rev. Mod. Phys.},
  number = {2}
}

@book{wilde2017a,
  title = {Quantum {{Information Theory}}},
  author = {Wilde, Mark M.},
  publisher = {{Cambridge University Press}},
  year = {2017},
  edition = {Second},
  isbn = {978-1-107-17616-4}
}

@article{wolf2006,
  title = {Extremality of {{Gaussian Quantum States}}},
  author = {Wolf, Michael M. and Giedke, Geza and Cirac, J. Ignacio},
  year = {2006},
  month = mar,
  volume = {96},
  pages = {080502},
  publisher = {{American Physical Society}},
  doi = {10.1103/PhysRevLett.96.080502},
  journal = {Phys. Rev. Lett.},
  number = {8}
}

@article{wu2010,
  title = {The Impact of Constellation Cardinality on Gaussian Channel Capacity},
  author = {Wu, Yihong and Verd{\'u}, Sergio},
  year = {2010},
  pages = {620--628},
  publisher = {{IEEE}},
  doi = {10.1109/ALLERTON.2010.5706965},
  isbn = {9781424482146},
  journal = {2010 48th Annual Allerton Conference on Communication, Control, and Computing, Allerton 2010}
}

@article{zhao2009,
  title = {Asymptotic Security of Binary Modulated Continuous-Variable Quantum Key Distribution under Collective Attacks},
  author = {Zhao, Yi Bo and Heid, Matthias and Rigas, Johannes and L{\"u}tkenhaus, Norbert},
  year = {2009},
  volume = {79},
  pages = {1--14},
  issn = {10502947},
  doi = {10.1103/PhysRevA.79.012307},
  journal = {Phys. Rev. A},
  number = {1}
}

@inproceedings{nguyen2004,
	location = {Parma, Italy, October},
	title = {Side-Information Coding with Turbo Codes and its Application to Quantum Key Distribution},
	booktitle = {International Symposium on Information Theory and its Applications},
	author = {{NGUYEN}, Kim-Chi and {VAN} {ASSCHE}, Gilles and {CERF}, Nicolas J},
	date = {2004-10-10},
	year = 2004
}

@inproceedings{bloch2006,
	location = {Punta del Este, Uruguay},
	title = {{LDPC}-based Gaussian key reconciliation},
	doi = {10.1109/ITW.2006.1633793},
	eventtitle = {2006 {IEEE} Information Theory Workshop},
	pages = {116--120},
	booktitle = {2006 {IEEE} Information Theory Workshop},
	publisher = {{IEEE}},
	author = {Bloch, M. and Thangaraj, A. and {McLaughlin}, S.W. and Merolla, J.-M.},
	year = {2006}
}

@incollection{holevo1973,
	title = {Statistical Problems in Quantum Physics},
	booktitle = {Proceedings of the {{Second Japan}}-{{USSR Symposium}} on {{Probability Theory}}},
	author = {Holevo, A. S.},
	editor = {Maruyama, G. and Prokhorov, Yu. V.},
	year = {1973},
	volume = {330},
	pages = {104--119},
	publisher = {{Springer Berlin Heidelberg}},
	address = {{Berlin, Heidelberg}},
	doi = {10.1007/BFb0061483}
}

\end{document}